\documentclass[preprint,prd,nofootinbib,tightenlines,groupedaddress,amsmath,amssymb]{revtex4}

\usepackage{bm}
\usepackage{color}
\usepackage{graphicx}
\usepackage[hypertex]{hyperref}
\usepackage{multirow}

\begin{document}
\baselineskip=15pt \parskip=3pt

\vspace*{3em}

\title{Higgs Decay \boldmath$h\to\mu\tau$ with Minimal Flavor Violation}

\author{Xiao-Gang He,$^{1,2,3}$ Jusak Tandean,$^{2}$ and Ya-Juan Zheng$^2$}
\affiliation{${}^{1}$INPAC, SKLPPC, and Department of Physics,
Shanghai Jiao Tong University, Shanghai 200240, China \vspace{3pt} \\
${}^{2}$CTS, CASTS, and
Department of Physics, National Taiwan University, Taipei 106, Taiwan \vspace{3pt} \\
${}^{3}$National Center for Theoretical Sciences and Physics Department of National
Tsing Hua University, Hsinchu 300, Taiwan \vspace{3ex}}


\begin{abstract}
We consider the tentative indication of flavor-violating Higgs boson decay $h\to\mu\tau$
recently reported in the CMS experiment within the framework of minimal flavor violation.
Specifically, we adopt the standard model extended with the seesaw mechanism involving
right-handed neutrinos plus effective dimension-six operators satisfying
the minimal flavor violation principle in the lepton sector.
We find that it is possible to accommodate the CMS \,$h\to\mu\tau$ signal interpretation
provided that the right-handed neutrinos couple to the Higgs boson in some nontrivial way.
We take into account empirical constraints from other lepton-flavor-violating processes
and discuss how future searches for the $\mu\to e\gamma$ decay and $\mu\to e$ conversion in
nuclei may further probe the lepton-flavor-violating Higgs couplings.
\end{abstract}

\maketitle

\section{Introduction\label{intro}}

The Higgs boson discovered at the LHC three years ago\,\,\cite{higgs@lhc} can offer
a potential window into physics beyond the standard model\,\,(SM).
The existence of new interactions can bring about modifications to the standard decay
modes of the particle and/or cause it to undergo exotic decays~\cite{Curtin:2013fra}.
As LHC data continues to accumulate with increasing precision, they may reveal
clues of new physics in the Higgs couplings.

The latest LHC measurements of the Higgs, $h$, have started to expose its Yukawa
interactions with leptons.
Particularly, the ATLAS and CMS Collaborations have observed the decay mode
\,$h\to\tau^+\tau^-$\, and measured its signal strength to be
\,$\sigma/\sigma_{\scriptscriptstyle\rm SM}^{}=1.44^{+0.42}_{-0.37}$
and $0.91\pm0.28$,\, respectively~\cite{atlas:h2tt,cms:h2tt}.
In contrast, their direct searches for the decay channel \,$h\to\mu^-\mu^+$\, have so far
come up with only upper limits on its branching fraction,
\,${\cal B}(h\to\mu^-\mu^+)<1.5\times10^{-3}$ and\,\,$1.6\times10^{-3}$,\,
respectively~\cite{atlas:h2mm,cms:h2mm}, at 95\% confidence level (CL).
Overall, these results are still consistent with SM expectations.

There have also been searches for flavor-violating dilepton Higgs decays, which the SM does
not accommodate.
In this regard, CMS recently reported\,\,\cite{cms:h2mt} the interesting detection of
a\,\,slight excess of \,$h\to\mu^\pm\tau^\mp$\, events with a\,\,significance of\,\,2.4$\sigma$.
If interpreted as a signal, the excess implies a branching fraction of
\,${\cal B}(h\to\mu\tau)={\cal B}(h\to\mu^-\tau^+)+{\cal B}(h\to\mu^+\tau^-)=
\bigl(0.84^{+0.39}_{-0.37}\bigr)$\%,\,
but as a statistical fluctuation it translates into the bound
\,${\cal B}(h\to\mu\tau)<1.51$\%\, at 95\% CL\,\,\cite{cms:h2mt}.
In view of its low statistical significance, it is too soon to draw a definite conclusion
from this finding, but it would constitute evidence of new physics if confirmed
by future experiments.

This tantalizing, albeit tentative, hint of lepton flavor violation (LFV) outside the neutrino
sector has attracted a growing amount of attention, as the detection of such a process would
serve as a test for many models\,\,\cite{h2ll',Dery:2013rta,Goudelis:2011un,Harnik:2012pb}
and could have major implications for upcoming Higgs
measurements\,\,\cite{Harnik:2012pb,colliders}.
Subsequent to the \,$h\to\mu\tau$\, announcement by CMS, its signal hypothesis was
theoretically examined in the contexts of various scenarios involving enlarged scalar
sectors\,\,\cite{h2mt,Dorsner:2015mja,Celis:2014roa,Lee:2014rba} or nonrenormalizable
effective interactions\,\,\cite{Dorsner:2015mja,Celis:2014roa,Lee:2014rba,Dery:2014kxa}.

In this paper, we follow the latter line of approach which relies on effective operators to
address LFV in Higgs decay.
To handle the LFV pattern systematically without getting into model details,
we adopt the framework of so-called minimal flavor violation (MFV).
Motivated by the fact that the SM has succeeded in describing the existing data
on flavor-changing neutral currents and $CP$ violation in the quark sector, the MFV principle
presupposes that Yukawa couplings are the only sources for the breaking of flavor and $CP$
symmetries~\cite{mfv1,D'Ambrosio:2002ex}.
However, unlike its straightforward implementation for quarks, there is no unique way
to extend the notion of MFV to leptons, as the minimal version of the SM by itself,
without right-handed neutrinos or extra scalar particles, does not accommodate LFV.
In light of the fact that flavor mixing among neutrinos has been empirically
established\,\,\cite{pdg}, it is attractive to formulate leptonic MFV by incorporating new
ingredients that can explain this observation~\cite{Cirigliano:2005ck}.
Thus, here we consider the SM expanded with the addition of three heavy right-handed neutrinos
as well as effective dimension-six operators conforming to the MFV criterion.\footnote{Various
scenarios of leptonic MFV have been discussed in
the literature\,\,\cite{Cirigliano:2005ck,Branco:2006hz,mlfv,He:2014fva,He:2014efa}.}
The heavy neutrinos are essential for the seesaw mechanism to endow light neutrinos with
Majorana masses.

In the next section, after briefly reviewing the MFV framework, we introduce
the effective dimension-six operators that can give rise to LFV in Higgs decay, only one of
which is relevant to\,\,\,$h\to\mu\tau$.\,
In Section\,\,\ref{numerics}, we explore the parameter space associated with this operator
which can yield \,${\cal B}(h\to\mu\tau)\sim1$\%,\, as CMS may have discovered.
At the same time, we take into account various experimental restrictions on the Higgs
couplings proceeding from the operator.
Specifically, we impose constraints inferred from the LHC measurements described above
as well as from the existing data on transitions with LFV that have long been the subject of
intensive quests, such as \,$\mu\to e\gamma$.\,
We present several sample points from the viable parameter space that can account for the CMS'
\,$h\to\mu\tau$\, signal interpretation.
We also discuss how future searches for \,$\mu\to e\gamma$\, and nuclear \,$\mu\to e$\,
conversion may offer further tests on the interactions of interest.
Finally, we look at a\,\,few other processes that can be induced by the same operator.
Especially, we find that the $Z$-boson decay \,$Z\to\mu\tau$\, can have a branching ratio
that is below its current empirical limit by merely less than an order of magnitude.
We make our conclusions in Section$\;$\ref{conclusion}.
An appendix contains some additional information and formulas.

\section{Operators with minimal lepton-flavor violaton\label{mfv}}

In the SM plus three right-handed Majorana neutrinos, the renormalizable Lagrangian for
lepton masses can be written as
\begin{eqnarray} \label{Lm}
{\mathcal L}_{\rm m}^{} \,\,=\,\,
- (Y_\nu)_{kl}^{}\,\overline{L}_{k,L\,}^{}\nu_{l,R\,}^{}\tilde H
- (Y_e)_{kl}^{}\,\overline{L}_{k,L\,}^{}E_{l,R\,}^{} H
\,-\, \tfrac{1}{2}\, (M_\nu)_{kl}^{}\,\overline{\nu^{\rm c}_{k,R}}\,\nu_{l,R}^{}
\;+\; {\rm H.c.} \;,
\end{eqnarray}
where \,$k,l=1,2,3$\, are implicitly summed over, $Y_{\nu,e}$ denote Yukawa coupling matrices,
$L_{k,L}$ stands for left-handed lepton doublets, $\nu_{l,R}^{}$ and $E_{l,R}$ represent
right-handed neutrinos and charged leptons, respectively, \,$\tilde H=i\tau_2^{}H^*$\, with
$\tau_2^{}$ being the second Pauli matrix and $H$ the Higgs doublet, $M_\nu$~is the Majorana
mass matrix of\,\,$\nu_{l,R}^{}$, and \,$\nu^{\rm c}_{k,R}\equiv(\nu_{k,R})^{\rm c}$,\,
the superscript referring to charge conjugation.
For the nonzero elements of $M_\nu$ taken to be much greater than those
of\,\,$v Y_\nu/\sqrt2$, the seesaw mechanism of type I is operational\,\,\cite{seesaw1} and
generates the light neutrinos' mass matrix
\,\mbox{$m_\nu^{}=-(v^2/2)\, Y_\nu^{}M_\nu^{-1}Y_\nu^{\rm T}=
U_{\scriptscriptstyle\rm PMNS\,}^{}\hat m_{\nu\,}^{}U_{\scriptscriptstyle\rm PMNS}^{\rm T}$},\,
where \,$v\simeq246$\,GeV\, is the Higgs's vacuum expectation value,
$U_{\scriptscriptstyle\rm PMNS}$ denotes the Pontecorvo-Maki-Nakagawa-Sakata
(PMNS~\cite{pmns}) matrix, and \,$\hat m_\nu^{}={\rm diag}\bigl(m_1^{},m_2^{},m_3^{}\bigr)$\,
contains the light neutrinos' eigenmasses.
This suggests\,\,\cite{Casas:2001sr}
\begin{eqnarray} \label{Ynum}
Y_\nu^{} \,\,=\,\,
\frac{i\sqrt2}{v}\,U_{\scriptscriptstyle\rm PMNS\,}^{}\hat m^{1/2}_\nu OM_\nu^{1/2}~,
\end{eqnarray}
where $O$ in general is a complex 3$\times$3 matrix satisfying \,$OO^{\rm T}=\openone$,\,
the right-hand side being a\,\,unit matrix, and can be parameterized as
\begin{eqnarray} \label{Omatrix}
O \,\,=\,\, e^{i\sf R}e^{{\sf R}'} \;, ~~~ ~~~~
{\sf R}^{(\prime)} \,\,=\, \left(\!\begin{array}{ccc}
0 & r_1^{\scriptscriptstyle(\prime)} & r_2^{\scriptscriptstyle(\prime)} \vspace{1pt} \\
-r_1^{\scriptscriptstyle(\prime)} & 0 & r_3^{\scriptscriptstyle(\prime)} \vspace{1pt} \\
-r_2^{\scriptscriptstyle(\prime)} & \mbox{$-r_3^{\scriptscriptstyle(\prime)}$} & 0
\end{array}\!\right)
\end{eqnarray}
with $r_{1,2,3}^{}$ and $r_{1,2,3}'$ being independent real constants.
Hence nonvanishing $r_{1,2,3}^{\scriptscriptstyle(\prime)}$ dictate how the Higgs
couples to the right-handed neutrinos in a nontrivial way according to Eq.\,(\ref{Ynum}).
Hereafter, we concentrate on the possibility that the right-handed neutrinos are degenerate,
so that \,$M_\nu={\mathcal M}\openone$.\,
In this particular scenario, only the $e^{i\sf R}$ part of $O$ matters
physically\,\,\cite{Branco:2006hz}.

The MFV hypothesis\,\,\cite{D'Ambrosio:2002ex,Cirigliano:2005ck} then implies that
${\mathcal L}_{\rm m}$ is formally invariant under the global flavor group
\,$G_\ell={\rm SU}(3)_L\times{\rm O}(3)_\nu\times{\rm SU}(3)_E$.\,
This entails that $L_{k,L}$, $\nu_{k,R}$, and $E_{k,R}$ belong to the fundamental
representations of their respective flavor groups,
\begin{eqnarray}
L_L^{} &\,\to\,&  V_L^{}L_L^{} \,, ~~~ ~~~~ \nu_R^{} \,\to\, {\mathcal O}_\nu^{}\nu_R^{} \,,
~~~ ~~~~ E_R^{} \,\to\, V_E^{}E_R^{} \,,
\end{eqnarray}
where \,$V_{L,E}\in{\rm SU}(3)_{L,E}$\, and \,${\mathcal O}_\nu\in{\rm O}(3)_\nu$\,
is an orthogonal real matrix\,\,\cite{D'Ambrosio:2002ex,Cirigliano:2005ck,Branco:2006hz}.
Furthermore, under $G_\ell$ the Yukawa couplings transform in the spurion sense
according to
\begin{eqnarray}
Y_\nu^{} \,\to\, V_L^{}Y_\nu^{}{\mathcal O}_\nu^{\rm T} \,, ~~~ ~~~~
Y_e^{} \,\to\, V_L^{}Y_e^{}V^\dagger_E \,.
\end{eqnarray}

Due to the symmetry under $G_\ell$, we can work in the basis where
\,$Y_e^{}=\sqrt2\;{\rm diag}\bigl(m_e^{},m_\mu^{},m_\tau^{}\bigr)/v$\,
and the fields $\tilde\nu_{k,L}$, $\nu_{k,R}$, and $E_k$ refer to
the mass eigenstates.
Explicitly, \,$(E_1,E_2,E_3)=(e,\mu,\tau)$.\,
We can then express $L_{k,L}$ in relation to $U_{\scriptscriptstyle\rm PMNS}$~as
\begin{eqnarray} \label{QL}
L_{k,L}^{} = \left(\!\begin{array}{c} (U_{\scriptscriptstyle\rm PMNS})_{kl\,}^{}
\tilde\nu_{l,L}^{} \vspace{2pt} \\ E_{k,L}^{} \end{array}\!\right) .
\end{eqnarray}
In the standard parametrization\,\,\cite{pdg}
\begin{eqnarray} \label{pmns}
U_{\scriptscriptstyle\rm PMNS}^{} \,= \left(\!\begin{array}{ccc}
 c_{12\,}^{}c_{13}^{} & s_{12\,}^{}c_{13}^{} & e^{-i\delta}s_{13}^{}
\vspace{1pt} \\
-s_{12\,}^{}c_{23}^{}-c_{12\,}^{}s_{23\,}^{}s_{13}^{}\,e^{i\delta} &
 c_{12\,}^{}c_{23}^{}-s_{12\,}^{}s_{23\,}^{}s_{13}^{}\,e^{i\delta} & s_{23\,}^{}c_{13}^{}
\vspace{1pt} \\
 s_{12\,}^{}s_{23}^{}-c_{12\,}^{}c_{23\,}^{}s_{13}^{}\,e^{i\delta} &
-c_{12\,}^{}s_{23}^{}-s_{12\,}^{}c_{23\,}^{}s_{13}^{}\,e^{i\delta} & c_{23\,}^{}c_{13}^{}
\end{array}\!\right) {\rm diag}\bigl(e^{i\alpha_1/2},e^{i\alpha_2/2},1\bigr) , ~~
\end{eqnarray}
where $\delta$ and $\alpha_{1,2}$ are the $CP$-violating Dirac and Majorana phases,
respectively, \,$c_{kl}^{}=\cos\theta_{kl}^{}$,\, and \,$s_{kl}^{}=\sin\theta_{kl}^{}$.

To put together effective Lagrangians beyond the SM with MFV built-in, one inserts products
of the Yukawa matrices among the pertinent fields to assemble $G_\ell$-invariant operators
that are singlet under the SM gauge group~\cite{D'Ambrosio:2002ex,Cirigliano:2005ck}.
Of interest here are the combinations
\begin{eqnarray} \label{ABl}
{\textsf A} \,\,=\,\, Y_\nu^{}Y_\nu^\dagger \,\,=\,\, \frac{2\mathcal M}{v^2}\,
U_{\scriptscriptstyle\rm PMNS\,}^{} \hat m^{1/2}_\nu O O^\dagger\hat m^{1/2}_\nu
U_{\scriptscriptstyle\rm PMNS}^\dagger \,, ~~~ ~~~~
{\textsf B} \,\,=\,\, Y_e^{}Y_e^\dagger \,\,=\,\,
{\rm diag}\bigl(y_e^2,y_\mu^2,y_\tau^2\bigr) \,,
\end{eqnarray}
where \,$y_f^{}=\sqrt 2\,m_f^{}/v$.\,
With these matrices, one can generally devise an object $\Delta$ as an infinite
power series in them and their products, but it turns out to be resummable into only 17
terms~\cite{Colangelo:2008qp}.
To maximize the new-physics effects, we assume that the right-handed neutrinos' mass $\mathcal M$
is large enough to render the biggest eigenvalue of $\textsf A$ equal to unity, which
conforms to the perturbativity requirement\,\,\cite{He:2014fva,Colangelo:2008qp}.
Given that the eigenvalues of $\textsf B$ are at most \,$y_\tau^2\sim1\times10^{-4}$,\, we
may consequently drop from $\Delta$ all the terms with $\textsf B$, which would otherwise
be needed in a study concerning $CP$ violation\,\,\cite{He:2014fva,He:2014efa}.
Accordingly, the relevant building block is\,\,\cite{He:2014efa}
\begin{eqnarray} \label{Delta}
\Delta \,\,=\,\, \xi_1^{}\openone + \xi_{2\,}^{}{\textsf A} + \xi_{4\,}^{}{\textsf A}^2 \,,
\end{eqnarray}
where in our model-independent approach $\xi_{1,2,4}^{}$ are free parameters expected to be
at most of\,\,${\mathcal O}(1)$, one or more of which could be suppressed or vanish,
depending on the underlying theory.
As \,${\rm Im}_{\,}\xi_{1,2,4}^{}$\, are tiny\,\,\cite{He:2014fva,Colangelo:2008qp},
we can further approximate \,$\Delta^\dagger=\Delta$.\,

One could then construct the desired $G_\ell$-invariant effective Lagrangians that are SM
gauge singlet.
The one pertaining to \,$h\to\ell\ell'$\, at tree level is given by\,\,\cite{Cirigliano:2005ck}
\begin{eqnarray} \label{Lmfv}
{\cal L}_{\rm MFV}^{} \,\,=\,\, \frac{O_{RL}^{(e3)}}{\Lambda^2} \;+\; {\rm H.c.} \,, ~~~~~~~
O_{RL}^{(e3)} \,\,=\,\, ({\cal D}^\rho H)^{\dagger\,}\bar E_R^{}Y_e^\dagger
\Delta_{\,}{\cal D}_\rho^{}L_L^{} \;,
\end{eqnarray}
where the mass scale $\Lambda$ characterizes the underlying heavy new-physics and
the covariant derivatives
\,${\cal D}^\rho H=\partial^\rho H+i\big(g_{\,}\tau_a^{}W_a^\rho+g' B^\rho\big)H/2$\, and
\,${\cal D}^\rho L=\partial^\rho L+i\big(g_{\,}\tau_a^{}W_a^\rho-g' B^\rho\big)L/2$\,
contain the usual SU(2)$_L\times{\rm U(1)}_Y$ gauge fields \,$W_a^\rho$\, and $B^\rho$
with coupling constants $g$ and~$g'$, respectively, and Pauli matrices $\tau_a^{}$, with
summation over \,$a=1,2,3$\, being implicit.
There are other dimension-six MFV operators involving $H$ and leptons that have been
written down\,\,\cite{Cirigliano:2005ck},
\begin{eqnarray} \label{others}
\begin{array}{ll}
i\bigl[H^{\dagger\,}{\cal D}_\rho H-({\cal D}_\rho H)^\dagger H \bigr]
\bar L_L^{}\gamma^\rho\Delta_{\scriptscriptstyle LL}^{}L_L^{} \;,  &
g'\bar E_R^{}Y^\dagger_e\Delta_{\scriptscriptstyle RL\,}^{}\sigma_{\rho\omega}^{}
H^\dagger L_L^{}B^{\rho\omega} \;,
\vspace{1ex} \\
i\bigl[ H^\dagger\tau_{a\,}^{}{\cal D}_\rho H
- ({\cal D}_\rho H)^\dagger\tau_a^{}H \bigr] \bar L_L^{}\gamma^\rho
\Delta_{\scriptscriptstyle LL}'\tau_a^{}L_L^{} \;, ~~~ ~~~~ &
g_{\,}\bar E_R^{}Y^\dagger_e\Delta_{\scriptscriptstyle RL\,}'\sigma_{\rho\omega}^{}
H^\dagger\tau_a^{}L_L^{}W_a^{\rho\omega} \;,
\end{array}
\end{eqnarray}
with $\Delta_{{\scriptscriptstyle LL},\scriptscriptstyle RL}^{\scriptscriptstyle(\prime)}$
being of the form of $\Delta$ in Eq.\,(\ref{Delta}) and having
their own coefficients $\xi_j$, but these operators do not induce tree-level dilepton
Higgs couplings.
The same thing can be said of the comparatively more suppressed
\,$i\bigl[H^{\dagger\,}{\cal D}_\rho H-({\cal D}_\rho H)^\dagger H \bigr]
\bar E_R\gamma^\rho Y_e^\dagger\Delta_{\scriptscriptstyle RR}Y_e E_R$.\,
In the literature the operator $H^\dagger H_{\,}\bar E_R Y_e^\dagger\Delta H^\dagger L_L$
is also often considered ({\it e.g.},\,\,\cite{Dery:2013rta}), but it can be shown to be
related to $O_{RL}^{(e3)}$ and the other operators above.
Explicitly, employing the equations of motions for SM fields \cite{Grzadkowski:2010es},
one can derive \cite{He:2014efa}
\begin{eqnarray} \label{relation}
O_{RL}^{(e3)}+{\rm H.c.} &=&
\frac{i}{8}\big[H^\dagger{\cal D}_\rho H-({\cal D}_\rho H)^\dagger H\big] \big(
\bar L_L^{}\gamma^\rho \big\{\Delta,Y_e^{}Y_e^\dagger\big\}L_L^{}
+ 4 \bar E_R^{}\gamma^\rho Y_e^\dagger\Delta Y_e^{}E_R^{} \big)
\nonumber \\ && +\;
\frac{i}{8}\big[H^\dagger\tau_a^{}{\cal D}_\rho H-({\cal D}_\rho H)^\dagger\tau_a^{}H\big]
\bar L_L^{}\gamma^\rho \big\{\Delta,Y_e^{}Y_e^\dagger\big\}\tau_a^{}L_L^{}
\nonumber \\ && +\;
\frac{i}{8}\big[H^\dagger{\cal D}_\rho H+({\cal D}_\rho H)^\dagger H\big]
\bar L_L^{}\gamma^\rho \big[\Delta,Y_e^{}Y_e^\dagger\big]L_L^{}
\nonumber \\ && +\;
\frac{i}{8}\big[H^\dagger\tau_a^{}{\cal D}_\rho H+({\cal D}_\rho H)^\dagger\tau_a^{}H\big]
\bar L_L^{}\gamma^\rho \big[\Delta,Y_e^{}Y_e^\dagger\big]\tau_a^{}L_L^{}
\nonumber \\ && +\;
\frac{1}{8}\big[\big(4H^\dagger H/v^2-2\big)m_{h\,}^2\bar E_R^{}Y_e^\dagger\Delta H^\dagger L_L^{}
+ 4\bar L_L^{}Y_e^{}E_R^{}\,\bar E_R^{}Y_e^\dagger\Delta L_L^{}
\nonumber \\ && ~~~ ~~~
+ \bar E_R^{}Y_e^\dagger\Delta_{\,}\sigma_{\rho\omega}^{}H^\dagger
\big(g'B^{\rho\omega}+g_{\,}\tau_a^{}W_a^{\rho\omega}\big)L_L^{}
\,+\, {\rm H.c.} \big]
\end{eqnarray}
plus terms involving quark fields and total derivatives.\footnote{The formula for
\,$O_{RL}^{(e3)}+{\rm H.c.}$\, in the footnote 1 of Ref.\,\cite{He:2014efa} has several terms
missing and the wrong sign in the dipole ($\sigma_{\rho\omega}$) part.
These errors have been corrected here in Eq.\,\,(\ref{relation}).}
The third and fourth lines of this equation, which have $\big[\Delta,Y_e^{}Y_e^\dagger\big]$,
also supply contributions to \,$h\to\ell\ell'$,\, but they correspond to small,
${\cal O}\big(m_{\ell,\ell'}^2/m_h^2\big)$, effects that will be ignored later
in Eq.\,\,(\ref{Ykl}).

\section{Decay amplitudes and numerical analysis\label{numerics}}

One can express the effective Lagrangian describing the Higgs decays
\,$h\to\ell^-\ell^{\prime+},\ell^{\prime-}\ell^+$\, for \,$\ell\neq\ell'$\, as
\begin{eqnarray}
{\cal L}_{h\ell\ell'}^{} \,\,=\,\, -{\cal Y}_{\ell\ell'}^{}\,\overline{\ell}P_R^{~\;}\ell'
- {\cal Y}_{\ell'\ell}^{}\,\overline{\ell'}P_R^{~\;}\ell \;+\; {\rm H.c.} \;,
\end{eqnarray}
where ${\cal Y}_{\ell\ell',\ell'\ell}^{}$ denote the Yukawa couplings,
which are in general complex.
Hence the combined rate of \,$h\to\ell^-\ell^{\prime+},\ell^{\prime-}\ell^+$\, is
\begin{eqnarray}
\Gamma_{h\to\ell\ell'}^{} \,\,=\,\, \Gamma_{h\to\ell\bar\ell'}+\Gamma_{h\to\bar\ell\ell'}
\,\,=\,\, \frac{m_h^{}}{8\pi} \big(|{\cal Y}_{\ell\ell'}^{}|^2 +
|{\cal Y}_{\ell'\ell}^{}|^2\big) \,,
\end{eqnarray}
where the lepton masses have been neglected compared to $m_h^{}$.
The flavor-conserving decay \,$h\to\ell^-\ell^+$\, has a rate of
\,$\Gamma_{h\to\ell\bar\ell}=m_h^{}|{\cal Y}_{\ell\ell}^{}|^2/(8\pi)$.\,

The MFV Lagrangian in Eq.\,(\ref{Lmfv}) contributes to both flavor-conserving and -violating
Higgs decays.
Including the SM part, we can write for \,$h\to E_k^-E_l^+$\,
\begin{eqnarray} \label{Ykl}
{\cal Y}_{E_kE_l}^{} \,\,=\,\, \delta_{kl}^{}\,{\cal Y}_{E_kE_k}^{\scriptscriptstyle\rm SM}
\,-\, \frac{m_{E_l}^{}m_h^2}{2\Lambda^2v}\,\Delta_{kl}^{} \,,
\end{eqnarray}
where \,${\cal Y}_{E_kE_k}^{\scriptscriptstyle\rm SM}=m_{E_k}/v$\, at tree level.
It follows that \,$|{\cal Y}_{\ell\ell'}^{}|\ll|{\cal Y}_{\ell'\ell}^{}|$\, for
\,$\ell\ell'=e\mu,e\tau,\mu\tau$\, and ${\cal Y}_{\ell\ell}^{}$ are real in our MFV scenario.

These couplings enter the amplitudes for a variety of lepton-flavor-violating processes,
such as~$\mu\to e\gamma$,\, via one- and two-loop diagrams.
Therefore, they are subject to the pertinent empirical
constraints\,\,\cite{Goudelis:2011un,Harnik:2012pb}, the most stringent of which we list here,
assuming that the impact of these loop contributions is not much reduced by other new-physics
effects.
As we sketch in Appendix\,\,\ref{m2eg}, the current bound
\,${\cal B}(\mu\to e\gamma)_{\rm exp}^{}<5.7\times10^{-13}$\, \cite{pdg} translates into
\begin{eqnarray} \label{m2egconstr}
\sqrt{\big|\big({\cal Y}_{\mu\mu}^{}+r_\mu^{}\big){\cal Y}_{\mu e}^{} +
9.19\,{\cal Y}_{\mu\tau\,}^{}{\cal Y}_{\tau e}^{}\big|\raisebox{1pt}{$^2$} +
\big|\big({\cal Y}_{\mu\mu}^{}+r_\mu^{}\big){\cal Y}_{e\mu}^{} +
9.19\,{\cal Y}_{e\tau\,}^{}{\cal Y}_{\tau\mu}^{}\big|\raisebox{1pt}{$^2$}}
\,\,<\,\, 5.1\times10^{-7} \;, ~~~~
\end{eqnarray}
where \,$r_\mu^{}=0.29$.\,
From \,${\cal B}(\tau\to e\gamma)_{\rm exp}^{}<3.3\times10^{-8}$~\cite{pdg},
one extracts\,\,\cite{Harnik:2012pb,Dery:2013rta}
\begin{eqnarray} \label{etconstr}
\big|{\cal Y}_{\tau\tau}^{}+r_\tau^{}\big|
\sqrt{\big|{\cal Y}_{\tau e}^{}\big|\raisebox{1pt}{$^2$}
+ \big|{\cal Y}_{e\tau}^{}\big|\raisebox{1pt}{$^2$}}
\,\,<\,\, 5.2\times10^{-4} \;,
\end{eqnarray}
where \,$r_\tau^{}=0.03$.\,
In these inequalities, we have put more than two different couplings together, as they are
generally affected by ${\cal L}_{\rm MFV}$ at the same time, and dropped smaller terms.
The aforementioned CMS \,$h\to\mu\tau$\, result under the no-signal assumption
implies\,\,\cite{cms:h2mt}
\begin{eqnarray} \label{mtconstr}
\sqrt{\big|{\cal Y}_{\tau\mu}^{}\big|\raisebox{1pt}{$^2$} +
\big|{\cal Y}_{\mu\tau}^{}\big|\raisebox{1pt}{$^2$}} \,\,<\,\, 3.6\times10^{-3} \;,
\end{eqnarray}
which is \mbox{\footnotesize\,$\sim$\,}4 times stronger than
the restraint\,\,\cite{Harnik:2012pb} inferred from
\,${\cal B}(\tau\to\mu\gamma)_{\rm exp}^{}<4.4\times10^{-8}$~\cite{pdg}
and encompasses the range
\begin{eqnarray} \label{Ymt}
2.0\times10^{-3} \,\,<\,\, \sqrt{\big|{\cal Y}_{\tau\mu}^{}\big|\raisebox{1pt}{$^2$} +
\big|{\cal Y}_{\mu\tau}^{}\big|\raisebox{1pt}{$^2$}} \,\,<\,\, 3.3\times10^{-3}
\end{eqnarray}
implied by \,${\cal B}(h\to\mu\tau)=\bigl(0.84^{+0.39}_{-0.37}\bigr)$\%\, in
the CMS signal hypothesis\,\,\cite{cms:h2mt}.

The information on \,$h\to\mu^+\mu^-,\tau^+\tau^-$\, recently acquired by
ATLAS\,\,\cite{atlas:h2mm,atlas:h2tt} and CMS\,\,\cite{cms:h2mm,cms:h2tt} is also useful
for restricting new physics in ${\cal Y}_{\mu\mu,\tau\tau}$.
From the data described in Section\,\,\ref{intro}, we may require
\begin{eqnarray} \label{mmttconstraints}
\big|{\cal Y}_{\mu\mu}^{}/{\cal Y}_{\mu\mu}^{\scriptscriptstyle\rm SM}\big|\raisebox{1pt}{$^2$}
\,\,<\,\, 6.5 \;, ~~~~~~~
0.7 \,\,<\,\,
\big|{\cal Y}_{\tau\tau}^{}/{\cal Y}_{\tau\tau}^{\scriptscriptstyle\rm SM}\big|\raisebox{1pt}{$^2$}
\,\,<\,\, 1.8 \;,
\end{eqnarray}
where \,${\cal Y}_{\mu\mu}^{\scriptscriptstyle\rm SM}=4.24\times10^{-4}$\, and
\,${\cal Y}_{\tau\tau}^{\scriptscriptstyle\rm SM}=7.19\times10^{-3}$\, in the SM from the
rates \,$\Gamma_{h\to\mu\bar\mu}^{\scriptscriptstyle\rm SM}=894$\,eV\, and
\,$\Gamma_{h\to\tau\bar\tau}^{\scriptscriptstyle\rm SM}=257$\,keV\, \cite{lhctwiki}
for \,$m_h^{}=125.1$\,GeV.\,
These numbers allow one to see from Eqs.\,\,(\ref{m2egconstr}) and\,\,(\ref{etconstr}),
where $r_\mu^{}$ and $r_\tau^{}$ represent the 2-loop
effects\,\,\cite{Harnik:2012pb,Dery:2013rta}, that the 2-loop contribution to
\,$\mu\to e\gamma$\, is dominant in constraining ${\cal Y}_{e\mu,\mu e}^{}$, whereas
the 1- and 2-loop effects on \,$\tau\to e\gamma$\, are roughly comparable.

We now attempt to attain \,$|{\cal Y}_{\mu\tau}^{}|\sim0.003$\, corresponding
to the CMS hint of \,$h\to\mu\tau$\, by scanning the coefficients $\xi_{1,2,4}^{}$ in
\,$\Delta=\xi^{}_1\openone+\xi^{}_{2\,}{\textsf A}+\xi^{}_{4\,}{\textsf A}^2$\, which enter
the Yukawa couplings according to\,\,Eq.\,(\ref{Ykl}) and consequently are subject to
the restrictions in Eqs.\,\,(\ref{m2egconstr})-(\ref{mtconstr}) and\,\,(\ref{mmttconstraints}).
Given that in our MFV scenario \,${\cal Y}_{\ell\ell'}\propto m_{\ell'}^{}$\, if
\,$\ell\neq\ell'$,\, from this point on we neglect ${\cal Y}_{\mu e,\tau e,\tau\mu}$
in comparison to ${\cal Y}_{e\mu,e\tau,\mu\tau}$, respectively.

Since $\sf A$ in Eq.\,(\ref{ABl}) can be realized in many different ways, we consider first
the possibility that the orthogonal $O$ matrix is real, in which case
\begin{eqnarray} \label{ArealO}
{\sf A} \,\,=\,\, Y_\nu^{}Y_\nu^\dagger \,\,=\,\,
\frac{2_{\,}\mathcal M}{v^2}\,U_{\scriptscriptstyle\rm PMNS\,}^{}
\hat m_{\nu\,}^{}U_{\scriptscriptstyle\rm PMNS}^\dagger
\end{eqnarray}
and the right-handed neutrinos' Yukawa coupling matrix in Eq.\,(\ref{Ynum})
simplifies to
\,$Y_\nu\propto U_{\scriptscriptstyle\rm PMNS\,}^{}\hat m^{1/2}_\nu$,\,
somewhat similar to its Dirac-neutrino counterpart\,\,\cite{He:2014efa}.
Although $U_{\scriptscriptstyle\rm PMNS}$ has dependence on the Majorana phases
$\alpha_{1,2}$, as in Eq.\,(\ref{pmns}), they drop out of Eq.\,(\ref{ArealO}).

To proceed numerically, we employ the central values of neutrino mixing parameters
from a\,\,recent fit to global neutrino data~\cite{nudata}.
Most of the numbers depend on whether light neutrino masses have a~normal hierarchy~(NH),
\,$m_1^{}<m_2^{}<m_3^{}$,\, or an inverted one (IH), \,$m_3^{}<m_1^{}<m_2^{}$.\,
Since experimental information on the absolute scale of $m_{1,2,3}^{}$ is still far from
precise~\cite{pdg}, for definiteness we select \,$m_1^{}=0$ $(m_3^{}=0)$\, in the NH (IH) case.

With the preceding choices, after exploring the $\xi_{1,2,4}^{}$ parameter space, we find that
$|{\cal Y}_{\mu\tau}^{}|$ can only reach somewhere in the range of \,(1-2)$\times10^{-4}$.\,
This is caused by the constraint in Eq.\,(\ref{m2egconstr}), without which the upper bound
\,$|{\cal Y}_{\mu\tau}^{}|<0.0036$\, could be easily saturated.
Thus, to reproduce the signal range in Eq.\,(\ref{Ymt}), the form of $\sf A$ in
Eq.\,(\ref{ArealO}) is not sufficient, and we instead need one with a less simple structure,
to which we pay our attention next.\footnote{A similar conclusion was drawn in
Ref.\,\cite{Dery:2014kxa} from a semi-quantitative investigation focusing on an MFV
contribution that corresponds to the $\xi_2^{}$ term in our study.}

A more promising possibility is that the $O$ matrix in Eq.\,(\ref{ABl}) is complex,
which leads to
\begin{eqnarray} \label{AcomplexO}
{\sf A} \,\,=\,\, Y_\nu^{}Y_\nu^\dagger \,\,=\,\,
\frac{2}{v^2}\,{\cal M}_{\,}U_{\scriptscriptstyle\rm PMNS\,}^{}
\hat m^{1/2}_\nu OO^\dagger\hat m^{1/2}_\nu U_{\scriptscriptstyle\rm PMNS}^\dagger \;.
\end{eqnarray}
As mentioned in the previous section, one can express \,$O=e^{i\sf R}e^{{\sf R}'}$\, with
real antisymmetric matrices $\sf R$ and ${\sf R}'$ defined in\,\,Eq.\,(\ref{Omatrix}).
Accordingly, we have
\begin{eqnarray} \label{oo+}
OO^\dagger \,\,=\,\, e^{2i\sf R} \,\,=\,\, \openone \,+\,
i{\sf R}\,\frac{\sinh(2\tilde r)}{\tilde r} \,-\,
2{\sf R}^2\,\frac{\sinh^2\!\tilde r}{\tilde r^2} \;, ~~~~ ~~~
\tilde r \,=\, \sqrt{r_1^2+r_2^2+r_3^2} \;,
\end{eqnarray}
and so nonzero $r_{1,2,3}$ can serve as extra free parameters that may allow us to
achieve the desired size of $|{\cal Y}_{\mu\tau}^{}|$.
This can indeed be realized, as illustrated by the examples collected in Table\,\,\ref{table}.
The flavor-violating Yukawa couplings quoted in the last three columns have followed from their
dependence on the elements of $\Delta$ determined using the listed sets of $\alpha_{1,2}^{}$,
$r_{1,2,3}$, and $\xi_{1,2,4}^{}/\Lambda^2$ numbers, along with the central values of neutrino
mixing parameters from Ref.\,\,\cite{nudata}, again with \,$m_1^{}=0$ $\big(m_3^{}=0\big)$\,
if the light neutrino masses have a normal (inverted) hierarchy.
The table includes a couple of instances with nonvanishing Majorana phases
$\alpha_{1,2}^{}$, which are not yet measured and affect $\sf A$, as $OO^\dagger$ in
Eq.\,(\ref{AcomplexO}) is not diagonal.

\begin{table}[t]
\begin{tabular}{|c|cccccccc|cccccc|} \hline
& \multirow{2}{*}{$\displaystyle\frac{\alpha_1^{}}{\pi}$} &
\multirow{2}{*}{$\displaystyle\frac{\alpha_2^{}}{\pi}$} &
\multirow{2}{*}{~$r_1^{}$} & \multirow{2}{*}{~$r_2^{}$} & \multirow{2}{*}{~$r_3^{}$} &
\mbox{\scriptsize$10^5\,\xi_1^{}/\Lambda^2$} & \mbox{\scriptsize$10^5\,\xi_2^{}/\Lambda^2$}
& \mbox{\scriptsize$10^5\,\xi_4^{}/\Lambda^2$}\,$\vphantom{|^{\int_|^|}}$ &
\multirow{2}{*}{\,$\displaystyle\frac{{\cal Y}_{ee}^{}}
{{\cal Y}_{ee}^{{}^{\scriptscriptstyle\rm SM}}}$\,} &
\multirow{2}{*}{$\displaystyle\frac{{\cal Y}_{\mu\mu}^{}}
{{\cal Y}_{\mu\mu}^{{}^{\scriptscriptstyle\rm SM}}}$} &
\multirow{2}{*}{\,$\displaystyle\frac{{\cal Y}_{\tau\tau}^{}}
{{\cal Y}_{\tau\tau}^{{}^{\scriptscriptstyle\rm SM}}}$\,}
& \multirow{2}{*}{$\displaystyle\frac{|{\cal Y}_{e\mu}^{}|}{10\raisebox{0.3pt}{$^{-6}$}}$}
& \multirow{2}{*}{$\displaystyle\frac{|{\cal Y}_{e\tau}^{}|}{10\raisebox{0.3pt}{$^{-4}$}}$}
& \multirow{2}{*}{\,$\displaystyle\frac{|{\cal Y}_{\mu\tau}^{}|}{10\raisebox{0.3pt}{$^{-3}$}}$\,}
\\ & & & & & & $(\mbox{\scriptsize${\rm GeV}^{-2}$})_{\vphantom{\int^|}}$ &
$(\mbox{\scriptsize${\rm GeV}^{-2}$})$ & $(\mbox{\scriptsize${\rm GeV}^{-2}$})$\, & & & & & &
\\ \hline\hline \multirow{3}{*}{NH} &
0 & 0 & ~ 0.81 & $-1.7$ & $-0.89\vphantom{|^{\int^|}}$ & $-6.3$~ & ~ 6.2~ & ~ 5.4~ &
1.5  & 1.2  & 0.89  & 1.7 & 0.3 & 3.1$\vphantom{|_{\int_|}}$
\\ &
0 & 0 & $-0.86$ & ~ 1.8 & $-0.92$ & $-7.1$~ & ~ 8.7~ & ~ 4.5~ &
1.6  & 1.2  & 0.87  & 2.0 & 0.4 & 3.5$\vphantom{|_{\int_|}}$
\\ &
0 & 0.23 & \,~ 0.74\, & \,$-0.80$\, & \,$-0.20$\, & ~ 4.9~ & $-6.7$~ & $-5.9$~ &
0.63 & 0.93 & 1.3   & 1.7 & 2.2 & 3.2$\vphantom{|_|^{}}$
\\
\hline \multirow{3}{*}{~IH~} &
0 & 0 & ~ 0.04 & ~ 0.63 & $-0.93\vphantom{|^{\int^|}}$ & $-7.9$~ & ~ 8.8~ & ~ 2.6~ &
1.5  & 1.2  & 1.1   & 2.1 & 2.8 & 3.2$\vphantom{|_{\int_|}}$
\\ &
0 & 0 & ~ 0.02 & $-0.75$ & ~ 1.1$\vphantom{|^{\int}}$ & $-5.7$~ & ~ 3.8~ & ~ 8.1~ &
1.4  & 1.1  & 0.90  & 2.4 & 1.3 & 3.3$\vphantom{|_{\int_|}}$
\\ &
\,0.79\, & 1.3 & $-0.61$ & $-0.79$ & ~ 1.4$\vphantom{|^{\int}}$ & $-5.3$~ & ~ 5.0~ & ~ 7.6~ &
1.4  & 1.0  & 0.84  & 1.2 & 0.4 & 3.5$\vphantom{|_|^{}}$
\\ \hline
\end{tabular}
\caption{Higgs-lepton Yukawa couplings corresponding to sample values of the Majorana
phases $\alpha_{1,2}$, the parameters $r_{1,2,3}$ of the complex $O$ matrix, and
the coefficients $\xi_{1,2,4}^{}$ in the MFV building block $\Delta$ which can yield
\,$|{\cal Y}_{\mu\tau}^{}|\mbox{\footnotesize\,$\gtrsim$\,}3\times10^{-3}$.\,
The calculation of the NH (IH) results also relies on the measured neutrino mixing
parameters in the case of normal (inverted) hierarchy of neutrino masses.\label{table}}
\end{table}

In the table, we also collect the corresponding flavor-conserving Yukawa couplings divided
by their SM predictions, including ${\cal Y}_{ee}^{}$ for completeness, with
\,${\cal Y}_{ee}^{\scriptscriptstyle\rm SM}=m_e^{}/v=2.08\times10^{-6}$.\,
It is obvious that ${\cal Y}_{\ell\ell}^{}$ can be altered sizeably with respect to their
SM values.
Therefore, measurements of \,$h\to\mu^+\mu^-,\tau^+\tau^-$\, with improved precision in
the future can offer complementary tests on the new contributions.

Based on our numerical exploration, there are a few more remarks we would like to make.
First, we have noticed that the viable parameter ranges in the NH case are broader than
their IH counterparts.
Second, in many trials we observe that
\,$|{\cal Y}_{e\tau}^{}|\mbox{\footnotesize\,$\lesssim$\,}0.1|{\cal Y}_{\mu\tau}^{}|$\,
for the hypothetical signal regions, as Table\,\,\ref{table} also shows.
This pattern has implications that may be checked empirically in the future.
Third, in the absence of either $\xi_2^{}$ or $\xi_4^{}$ the maximal $|{\cal Y}_{\mu\tau}^{}|$
is somewhat lower than that when $\xi_{1,2,4}^{}$ are all contributing,
but at least some or all of the signal values in Eq.\,(\ref{Ymt}) can be accommodated.
However, if only $\xi_2^{}$, $\xi_4^{}$, or $\xi_{2,4}^{}$ are nonzero,
$|{\cal Y}_{\mu\tau}^{}|$ cannot exceed \mbox{\footnotesize\,$\sim$\,}0.0018.

Now, the six sample sets of parameter values in Table\,\,\ref{table} produce
branching fractions of \,$\mu\to e\gamma$ and $\tau\to\mu\gamma$\, in the ranges
of \,(1.4-5.4)$\times10^{-13}$\, and \,(1.6-2.0)$\times10^{-9}$,\, respectively,
if other new-physics effects are negligible.
The former numbers are within only a few times below the present bound
${\cal B}(\mu\to e\gamma)_{\rm exp}^{}$, whereas the latter are at least a factor of
20 less than ${\cal B}(\tau\to\mu\gamma)_{\rm exp}^{}$.
They can be regarded as predictions testable by ongoing or future experiments looking
for these decays if the CMS' indication of \,$h\to\mu\tau$\, is substantiated
by upcoming Higgs measurements and the signal range in Eq.\,(\ref{Ymt}), or part of it,
persists with increased data.
Especially, the planned MEG\,II experiment on \,$\mu\to e\gamma$,\, with sensitivity
expected to reach a few times $10^{-14}$ after 3 years of data
taking\,\,\cite{Cavoto:2014qoa,CeiA:2014wea}, will probe the above predictions for it.

As it turns out, if the forthcoming search for \,$\mu\to e\gamma$\, still comes up empty,
there could yet remain viable, but narrower, signal parameter regions.
We illustrate this in Table\,\,\ref{future}, assuming a\,\,possible future limit of
\,${\cal B}(\mu\to e\gamma)<5\times10^{-14}$\, \cite{Cavoto:2014qoa},
which amounts to replacing the right-hand side of Eq.\,(\ref{m2egconstr}) with
\,$1.5\times10^{-7}$,\, and also imposing the ratios
\,$0.5<\Gamma_{h\to\mu\bar\mu}/\Gamma_{h\to\mu\bar\mu}^{\scriptscriptstyle\rm SM}<1.5$\, and
\,$0.8<\Gamma_{h\to\tau\bar\tau}/\Gamma_{h\to\tau\bar\tau}^{\scriptscriptstyle\rm SM}<1.2$\,
based on LHC Run-2 projections\,\,\cite{lhcprojections}.
Since the examples in Table\,\,\ref{future} yield
\,${\cal B}(\mu\to e\gamma)=(1.2$-$4.4)\times10^{-14}$,\, they may be out of reach of MEG\,II,
and so to probe them one will likely need to rely on experiments looking for nuclear
\,$\mu\to e$\, conversion, which promise a greater degree of sensitivity in the long
run\,\,\cite{CeiA:2014wea}.
As discussed in Appendix\,\,\ref{m2eg}, the existing data on \,$\mu\to e$\, conversion in nuclei
are not yet competitive to the current measured bound on \,$\mu\to e\gamma$\, in constraining
the Yukawa couplings.
However, we also point out in the appendix that planned searches for \,$\mu\to e$\, conversion,
such as Mu2E and COMET\,\,\cite{CeiA:2014wea}, can be expected to test very well the parameter
space represented by the examples in Tables\,\,\ref{table} and\,\,\ref{future}.

\begin{table}[t]
\begin{tabular}{|c|cccccccc|cccccc|} \hline
& \multirow{2}{*}{$\displaystyle\frac{\alpha_1^{}}{\pi}$} &
\multirow{2}{*}{$\displaystyle\frac{\alpha_2^{}}{\pi}$} &
\multirow{2}{*}{~ $r_1^{}$} & \multirow{2}{*}{~ $r_2^{}$} & \multirow{2}{*}{~ $r_3^{}$} &
\mbox{\scriptsize$10^5\,\xi_1^{}/\Lambda^2$} & \mbox{\scriptsize$10^5\,\xi_2^{}/\Lambda^2$}
& \mbox{\scriptsize$10^5\,\xi_4^{}/\Lambda^2$}\,$\vphantom{|^{\int_|^|}}$ &
\multirow{2}{*}{\,$\displaystyle\frac{{\cal Y}_{ee}^{}}
{{\cal Y}_{ee}^{{}^{\scriptscriptstyle\rm SM}}}$\,} &
\multirow{2}{*}{$\displaystyle\frac{{\cal Y}_{\mu\mu}^{}}
{{\cal Y}_{\mu\mu}^{{}^{\scriptscriptstyle\rm SM}}}$} &
\multirow{2}{*}{\,$\displaystyle\frac{{\cal Y}_{\tau\tau}^{}}
{{\cal Y}_{\tau\tau}^{{}^{\scriptscriptstyle\rm SM}}}$\,}
& \multirow{2}{*}{$\displaystyle\frac{|{\cal Y}_{e\mu}^{}|}{10\raisebox{0.3pt}{$^{-6}$}}$}
& \multirow{2}{*}{$\displaystyle\frac{|{\cal Y}_{e\tau}^{}|}{10\raisebox{0.3pt}{$^{-4}$}}$}
& \multirow{2}{*}{\,$\displaystyle\frac{|{\cal Y}_{\mu\tau}^{}|}{10\raisebox{0.3pt}{$^{-3}$}}$\,}
\\ & & & & & & $(\mbox{\scriptsize${\rm GeV}^{-2}$})_{\vphantom{\int^|}}$ &
$(\mbox{\scriptsize${\rm GeV}^{-2}$})$ & $(\mbox{\scriptsize${\rm GeV}^{-2}$})$\, & & & & & &
\\ \hline\hline \multirow{2}{*}{NH} &
0 & 0 & $-0.53$ & ~ 0.73 & $-0.40\vphantom{|^{\int^|}}$ & ~ 6.0~ & $-0.7$~ & $-9.5$~ &
0.53 & 0.79 & 1.1   & 0.6 & 0.2 & 2.7$\vphantom{|_{\int_|}}$
\\ &
0 & \,0.4\, & \,~ 0.68\, & \,$-0.80$\, & \,$-0.15$\,   &  $-5.4$~  & $-2.3$~ & ~ 12~ &
1.4 & 1.2 & 0.93   & 0.3 & 0.5 & 2.6$\vphantom{|_|^{}}$
\\
\hline \multirow{2}{*}{~IH~} &
0 & 0 & ~ 0.0  & $-0.73$ & ~ 1.1$\vphantom{|^{\int^|}}$ & $-4.7$~ & $-1.9$~ & ~ 11~ &
1.4  & 1.1  & 0.96 & 0.5 & 0.1 & 2.5$\vphantom{|_{\int_|}}$
\\ &
~0.8~ & 1.3 & $-0.60$ & $-0.81$ & ~ 1.4$\vphantom{|^{\int}}$ & $-6.5$~ & ~ 9.4~ & ~ 1.1~ &
1.5  & 1.2  & 1.0  & 0.1 & 0.5 & 2.9$\vphantom{|_|^{}}$
\\ \hline
\end{tabular}
\caption{The same as Table\,\,\ref{table}, except the \,$\mu\to e\gamma$\, and
\,$h\to\mu\bar\mu,\tau\bar\tau$\, constraints are replaced with their projected
future experimental limits, as described in the text.\label{future}}
\end{table}

Finally, we discuss the contributions of ${\cal L}_{\rm MFV}$ in Eq.\,(\ref{Lmfv}) to
some other processes.
Expanding the operator, we have
\begin{eqnarray} \label{ORL}
O_{RL}^{(e3)} &=& \frac{\Delta_{kl\,}^{}m_{E_k}^{}}{v}\, \bar E_k^{}P_L^{}
\Bigg( \partial_\eta^{}E_l^{} - i e A_\eta^{}E_l^{} + i g_L^{}Z_\eta^{}E_l^{}
+ \frac{i g}{\sqrt2}\,W^-_\eta\nu_l^{} \Bigg) \partial^\eta h
\nonumber \\ && \! +\;
\frac{\Delta_{kl\,}^{}g_{\;\!}m_{E_k}^{}}{v}_{\,}\bar E_k^{}P_L^{} \Bigg[
\frac{i Z^{\eta\,}\partial_\eta E_l^{}}{2c_{\rm w}^{}} -
\frac{i W_\eta^{-\,}\partial^\eta\nu_l^{}}{\sqrt2}
+ \!\Bigg( \frac{e A\!\cdot\!Z}{2c_{\rm w}^{}} - \frac{g_L^{}Z^2}{2c_{\rm w}^{}}
+ \frac{g}{2}_{\,} W^+\!\!\cdot\!W^- \Bigg) E_l^{} \Bigg] (h+v) \,, \nonumber \\
\end{eqnarray}
where \,$g_L^{}=g\bigl(s_{\rm w}^2-1/2\bigr)/c_{\rm w}^{}$\, and
\,$c_{\rm w}^{}=\sqrt{1-s_{\rm w}^2}=g v/(2m_Z^{})=m_W^{}/m_Z^{}$.\,
Evidently, ${\cal L}_{\rm MFV}$ not only induces the already addressed \,$h\to\ell\bar\ell'$\,
couplings, but also contributes to the two-body decays of the weak bosons,
\,$Z\to\ell\bar\ell'$\, and \,$W\to\tau\nu_l^{}$,\,
as well as to three- and four-body modes, such as
\,$h\to\ell\bar\ell'\gamma,\nu\ell W^+,\ell\bar\ell'\gamma Z$.\,
Since the latter are more suppressed by phase space, we deal with only the two-body
$Z$ and $W$ decays.
The other operators in Eq.\,(\ref{others}) can also affect \,$Z\to\ell\bar\ell'$\, and
\,$W\to\tau\nu_l^{}$,\, but here we entertain the possibility that their impact is
comparatively unimportant.
Accordingly, from Eq.\,(\ref{ORL}) we derive
\begin{eqnarray} \label{MZ2ll'}
{\cal M}_{Z\to E_k\bar E_l}^{} &=&
\bar u_{E_k}^{}\bigg[ \delta_{kl}^{}\,\slash{\!\!\!\varepsilon}_Z^{}\,\big(g_L^{}P_L^{}+g_R^{}P_R^{}\big)
+ \frac{\Delta_{kl\,}^{}m_Z^{}}{\Lambda^{2\,}v}\Big( m_{E_k}^{}P_L^{}\,\varepsilon_Z^{}\cdot p_{E_l}^{}
- m_{E_l}^{}P_R^{}\,\varepsilon_Z^{}\cdot p_{E_k}^{} \Big) \bigg] v_{E_l}^{} \;, ~~~
\nonumber \\
{\cal M}_{W\to\tau\nu_l^{}}^{} &\,=\,& \bar u_\tau^{}\bigg( \frac{\delta_{3l}^{}\,g}{\sqrt2}\;
\slash{\!\!\!\varepsilon}_W^{} +
\frac{\sqrt2\,\Delta_{3l\,}^{}m_\tau^{}m_W^{}}{\Lambda^{2\,}v}\,
\varepsilon_W^{}\cdot p_\tau^{} \bigg) P_L^{}v_{\nu_l^{}}^{} \;.
\end{eqnarray}
where \,$g_R^{}=g\,s_{\rm w}^2/c_{\rm w}^{}$\, and we have included the SM terms in
these amplitudes.
Hence, neglecting lepton masses compared to $m_Z^{}$, we arrive at
\begin{eqnarray}
\Gamma_{Z\to\mu\bar e}^{} \,\,=\,\, \Gamma_{Z\to\mu\bar e}^{} \,\,\simeq\,\,
\frac{\big|\Delta_{12\,}^{}m_\mu^{}\big|^2m_Z^5}{192_{\,}\Lambda^{4\,}\pi_{\,}v^2}
\,\,=\,\, \frac{\big|{\cal Y}_{e\mu}^{}\big|^2 m_Z^5}{48\pi_{\,}m_h^4}
\end{eqnarray}
and similarly for \,$Z\to e\tau,\mu\tau$.\,
Thus, for, say, \,$\big|{\cal Y}_{e\mu}^{}\big|=2.1\times10^{-6}$,\,
$\big|{\cal Y}_{e\tau}^{}\big|=2.8\times10^{-4}$,\, and
\,$\big|{\cal Y}_{\mu\tau}^{}\big|=0.0032$\, from Table\,\,\ref{table}, we get
\begin{eqnarray}
{\cal B}\bigl(Z\to e^\pm\mu^\mp\bigr) = 6.0\times10^{-13} , ~~
{\cal B}\bigl(Z\to e^\pm\tau^\mp\bigr) = 1.1\times10^{-8} , ~~
{\cal B}\bigl(Z\to\mu^\pm\tau^\mp\bigr) = 1.4\times10^{-6} . ~~~
\end{eqnarray}
For comparison, the experimental limits are\,\,\cite{pdg}
\begin{eqnarray}
{\cal B}\bigl(Z\to e^\pm\mu^\mp\bigr)_{\rm exp} &<& 1.7\times10^{-6} \;, ~~~~ ~~~
{\cal B}\bigl(Z\to e^\pm\tau^\mp\bigr)_{\rm exp} \,<\, 9.8\times10^{-6} \;, ~~~~ ~~~
\nonumber \\ {\cal B}\bigl(Z\to\mu^\pm\tau^\mp\bigr)_{\rm exp} &<& 1.2\times10^{-5}
\end{eqnarray}
at 95\% CL.
We see that the predicted ${\cal B}(Z\to\mu\tau)$ is below its experimental bound by only
less than a\,\,factor of 10.
Therefore, \,$Z\to\mu\tau$\, is potentially more testable than \,$Z\to e\mu,e\tau$,\, and
the quest for it can provide a complementary check on ${\cal L}_{\rm MFV}$.

Neglecting lepton masses compared to $m_{W,Z}^{}$, we also obtain from Eq.\,(\ref{MZ2ll'})
\begin{eqnarray}
\Gamma_{Z\to E_k\bar E_k}^{} &\,=\,& \frac{m_Z^{}}{24\pi} \Bigg( g_L^2+g_R^2 +
\frac{\Delta_{kk\,}^2m_{E_k}^2m_Z^4}{4\Lambda^{4\,}v^2} \Bigg) \,,
\nonumber \\
\Gamma_{W\to\tau\nu}^{} &\,=\,& \frac{m_W^{}}{48\pi} \Bigg( g^2 +
\frac{\Delta_{33\,}^2m_{\tau\,}^2m_W^4}{2\Lambda^{4\,}v^2} \Bigg) +
\frac{\big(|\Delta_{31}|^2+|\Delta_{32}|^2\big)m_{\tau\,}^2m_W^4}{96\Lambda^{4\,}\pi_{\,}v^2} \;,
\end{eqnarray}
where in the \,$W\to\tau\nu$\, formula we have summed over the 3 neutrino flavors.
For the parameter values in Table\,\,\ref{table}, the nonstandard terms in
$\Gamma_{Z\to E_k\bar E_k}$ and $\Gamma_{W\to\tau\nu}^{}$ are tiny, being smaller than
the SM parts by more than 4 orders of magnitude.

Before ending this section, we would like to note that all the preceding analysis can be
repeated within the context of the type-III seesaw model \cite{seesaw3} with MFV, which is
very similar to the type-I case addressed in this study if the triplet leptons in the former
are as heavy as the right-handed neutrinos in the latter \cite{He:2014efa}.
However, in the type-II seesaw model \cite{seesaw2} with MFV, the Yukawa coupling matrix of
the triplet scalars does not possess the special feature that $Y_\nu$ has with regard to
the $O$ matrix\,\,\cite{He:2014efa} that allows ${\cal Y}_{\mu\tau}^{}$ to become large
enough to explain the CMS \,$h\to\mu\tau$\, signal hypothesis.

\section{Conclusions\label{conclusion}}

We have explored the possibility that the slight excess of \,$h\to\mu\tau$\, events recently
detected in the CMS experiment has a new-physics origin.
Adopting in particular the effective theory framework of MFV, we consider the SM extended
with the type-I seesaw mechanism and an effective dimension-six operator responsible for
the flavor-violating dilepton Higgs decay.
We demonstrate that to account for the tentative \,$h\to\mu\tau$\, signal, with a branching
fraction of order\,\,1\%, the Yukawa coupling matrix of the right-handed neutrinos needs to
have a nontrivial structure because of the stringent empirical constraints.
To illustrate this, we present several benchmark points that have survived the restrictions
from the existing \,$\mu\to e\gamma$,\, $\tau\to e\gamma$,\, and
\,$h\to\mu\bar\mu,\tau\bar\tau$\, data.
The viable parameter space can be probed further by upcoming LHC measurements and future
quests for charged-lepton-flavor violation.
Lastly, we examine a few other transitions that arise from the same dimension-six operator,
among which \,$Z\to\mu\tau$\, can have a predicted branching ratio merely less than 10 times
below its current empirical limit and hence potentially also testable in near-future searches.

\acknowledgments

This research was supported in part by the MOE Academic Excellence Program (Grant No. 102R891505)
and NSC of ROC and by NSFC (Grant No. 11175115) and Shanghai Science and Technology Commission
(Grant No. 11DZ2260700) of PRC.

\appendix

\section{Constraints from \boldmath$\mu\to e\gamma$ decay and $\mu\to e$ conversion\label{m2eg}}

The effective Lagrangian for \,$\mu\to e\gamma$\, can be expressed as
\begin{eqnarray}
{\cal L}_{\mu\to e\gamma}^{} \,\,=\,\, \frac{\sqrt{\alpha\pi}\,m_\mu^{}}{4\pi^2}\,
\overline{e}\,\sigma^{\rho\omega}\big({\cal C}_L^{}P_L^{}
+ {\cal C}_R^{}P_R^{}\big)\mu\,F_{\rho\omega}^{} \;,
\end{eqnarray}
where \,$\alpha=1/137$\, is the fine structure constant, \,$P_{L,R}=(1\mp\gamma_5)/2$,\,
and $F_{\rho\omega}$ is the electromagnetic field strength tensor.
This leads to the decay rate
\begin{eqnarray}
\Gamma_{\mu\to e\gamma}^{} \,\,=\,\, \frac{\alpha_{\;\!}m_\mu^5}{64\pi^4} \big(
|{\cal C}_L|^2+|{\cal C}_R|^2\big) \;,
\end{eqnarray}
The Wilson coefficients ${\cal C}_{L,R}$ receive contributions from Higgs-mediated
one-loop and two-loop\,\,\cite{Chang:1993kw} diagrams,
\,${\cal C}_{L,R}^{}={\cal C}_{L,R}^{\rm 1\,loop} + {\cal C}_{L,R}^{\rm 2\,loop}$.\,
Given that ${\cal Y}_{\ell\ell}^{}$ is real and
\,$|{\cal Y}_{ee}^{}|\ll|{\cal Y}_{\mu\mu}^{}|$,\, one finds\,\,\cite{Harnik:2012pb}
\begin{eqnarray}
{\cal C}_R^{\rm 1\,loop} &\simeq&
\frac{{\cal Y}_{\mu\mu\,}^{}{\cal Y}_{e\mu}^{}}{2 m_h^2} \bigg(
\!\log\frac{m_h^{}}{m_\mu^{}}-\frac{2}{3}\bigg) +
\frac{m_{\tau\,}^{}{\cal Y}_{e\tau\,}^{}{\cal Y}_{\tau\mu}^{}}{2 m_{\mu\,}^{}m_h^2} \bigg(
\!\log\frac{m_h^{}}{m_\tau^{}}-\frac{3}{4}\bigg) \,,
\nonumber \\
{\cal C}_R^{\rm 2\,loop} &\simeq&
\frac{0.055\,m_{\tau\,}^{}{\cal Y}_{e\mu}^{}}{m_{\mu\,}^{}m_h^2}
\end{eqnarray}
and ${\cal C}_L^{\rm1\,loop,2\,loop}$ obtainable from ${\cal C}_R^{\rm1\,loop,2\,loop}$  with
the replacements \,${\cal Y}_{\ell\ell'}^{}\to{\cal Y}_{\ell'\ell}^*$.\,
Here we suppose that there are no other new-physics contributions that can bring about
destructive interference with these coefficients.
Thus, putting together these formulas with the latest experimental bound\,\,\cite{pdg}
\,${\cal B}(\mu\to e\gamma)_{\rm exp}^{}<5.7\times10^{-13}$,\,
we arrive at Eq.\,(\ref{m2egconstr}) for \,$m_h^{}=125.1$\,GeV,\, which is consistent
with the most recent measurement\,\,\cite{lhc:mh}.

The effective Lagrangian for \,$\mu\to e$\, conversion in nuclei is \cite{Kitano:2002mt}
\begin{eqnarray}
{\cal L}_{\mu\to e}^{} \,=\, \frac{\sqrt{\alpha\pi}\,m_\mu^{}}{4\pi^2}\,
\overline{e}\,\sigma^{\rho\omega}\big({\cal C}_L^{}P_L^{}
+ {\cal C}_R^{}P_R^{}\big)\mu\,F_{\rho\omega}^{}
- \frac{1}{2}~\raisebox{-7pt}{\Large$\stackrel{\sum}{\mbox{\scriptsize$q$}}$}~
\overline{e}\big(g_{LS}^q P_R^{}+g_{RS}^q P_L^{}\big)\mu\,\bar q q \;,
\end{eqnarray}
where $q$ runs over all quark flavors, we have displayed only the most important terms for
our purposes, and, if ${\cal Y}_{\ell\ell'}$ are the only LFV sources, ${\cal C}_{L,R}$ are
already written down in the preceding paragraph and\,\,\cite{Harnik:2012pb}
\begin{eqnarray}
g_{LS}^q \,\,=\,\, \frac{-2 m_{q\,}^{}{\cal Y}_{\mu e}^*}{m_h^{2\,}v} \;, ~~~ ~~~~
g_{RS}^q \,\,=\,\, \frac{-2 m_{q\,}^{}{\cal Y}_{e\mu}^{}}{m_h^{2\,}v} \;.
\end{eqnarray}
The \,$\mu\to e$\, conversion rate in nucleus $\cal N$ is then given by~\cite{Kitano:2002mt}
\begin{eqnarray} \label{Bm2e}
{\cal B}(\mu {\cal N}\to e{\cal N}) &=& \frac{m_\mu^5}{\omega_{\rm capt}^{\cal N}}
\bigg|\frac{\sqrt{\alpha\pi}~{\cal C}_{L\,}^{}D_{\cal N}^{}}{8\pi^2} \,-\,
\tilde g_{LS\,}^{(p)}S_{\cal N}^{(p)} \,-\,
\tilde g_{LS\,}^{(n)}S_{\cal N}^{(n)}\bigg|_{\vphantom{|_|^|}}\raisebox{9pt}{$^2$}
\;+\; (L\to R) \;, \\
\tilde g_{LS\,}^{(N)} \,=\, \raisebox{-7pt}{\Large$\stackrel{\sum}{\mbox{\scriptsize$q$}}$}\,
\frac{g_{LS}^q}{m_q^{}}f_q^{(N)}m_N^{}
&=& \frac{-2m_{N}^{}{\cal Y}_{\mu e}^*}{m_h^{2\,}v}\,
\raisebox{-7pt}{\Large$\stackrel{\sum}{\mbox{\scriptsize$q$}}$}\,f_q^{(N)} \,, ~~~~
f_q^{(N)} = \frac{\langle N|m_{q\,}^{}\bar q q|N\rangle}{m_N^{}} \,, ~~~~
N = p,n \,, ~~~~~
\end{eqnarray}
where $D_{\cal N}^{}$ and $S_{\cal N}^{(p,n)}$ are dimensionless integrals
representing the overlap of electron and muon wave functions for $\cal N$ and
$\omega_{\rm capt}^{\cal N}$ is the rate of muon capture in $\cal N$.
Based on the current experimental limits on \,$\mu\to e$\, transition in various
nuclei~\cite{pdg,Papoulias:2013gha} and the corresponding overlap integral and
$\omega_{\rm capt}^{\cal N}$ values~\cite{Kitano:2002mt}, one expects that
the \,${\cal N}=\rm Au$ and Ti\, data may supply the most consequential restrictions.
The evaluation of ${\cal B}(\mu {\cal N}\to e{\cal N})$ for these two nuclei,
respectively, requires \,$D_{\rm Au}=0.189$,\, $D_{\rm Ti}=0.087$,\,
$S_{\rm Au}^{(p)}=0.0614$,\, $S_{\rm Au}^{(n)}=0.0918$,\,
$S_{\rm Ti}^{(p)}=0.0368$,\, $S_{\rm Ti}^{(n)}=0.0435$,\,
$\omega_{\rm capt}^{\rm Au}=13.07\times10^6/\rm s$,\, and
\,$\omega_{\rm capt}^{\rm Ti}=2.59\times10^6/\rm s$\, \cite{Kitano:2002mt},
as well as the latest determination of the sum of the nucleon matrix elements,
\,$\Sigma_q f_q^{(p,n)}=0.305\pm0.009$\, \cite{Hoferichter:2015dsa},\footnote{This sum
partly depends on the pion-nucleon $\sigma$-term determined in
Ref.\,\cite{Hoferichter:2015dsa} which agrees with that
previously calculated in Ref.\,\cite{Alarcon:2011zs}.}
which lies around the lower end of the ranges from some of earlier
estimates\,\,\cite{He:2008qm}.

If we impose the measured bound
\,${\cal B}(\mu_{\,\!}{\rm Au}\to e_{\,\!}{\rm Au})_{\rm exp}<7\times10^{-13}$~\cite{pdg},
instead of Eq.\,(\ref{m2egconstr}), but still apply Eqs.\,\,(\ref{etconstr}), (\ref{mtconstr}),
and\,\,(\ref{mmttconstraints}), we end up with \,$|{\cal Y}_{e\mu}^{}|<1.6\times10^{-5}$,\,
which is compatible with the finding of Ref.\,\cite{Crivellin:2014cta}.
If we use \,${\cal N}=\rm Ti$\, with
\,${\cal B}(\mu_{\,\!}{\rm Ti}\to e_{\,\!}{\rm Ti})_{\rm exp}
<6.1\times10^{-13}$~\cite{Papoulias:2013gha},
instead of \,${\cal N}=\rm Au$,\, we get the somewhat stricter
\,$|{\cal Y}_{e\mu}^{}|<1.3\times10^{-5}$.\,
These limitations are roughly 5 to 13 times higher than the range of results
\,$|{\cal Y}_{e\mu}^{}|=($1.2-2.4$)\times10^{-6}$\, quoted in
Table\,\,\ref{table}, demonstrating that the present data on nuclear \,$\mu\to e$\,
conversion are not yet competitive to \,${\cal B}(\mu\to e\gamma)_{\rm exp}$\,
in restricting especially ${\cal Y}_{e\mu}^{}$, which is also known in
the literature\,\,\cite{Goudelis:2011un,Harnik:2012pb,Crivellin:2014cta}.
Nevertheless, the leading planned searches for \,$\mu\to e$\, conversion,
Mu2E and COMET, which utilize aluminum as the target material\,\,\cite{CeiA:2014wea},
will likely be able to probe the parameter space represented by the examples in both
Tables\,\,\ref{table} and\,\,\ref{future}.
More precisely, from the sets of sample numbers in these tables, together with
the aluminum parameters \,$D_{\rm Al}=0.0362$,\, $S_{\rm Al}^{(p)}=0.0155$,\,
$S_{\rm Al}^{(n)}=0.0167$,\, and
$\omega_{\rm capt}^{\rm Al}=0.7054\times10^6/\rm s$\, \cite{Kitano:2002mt}, we obtain
\,${\cal B}(\mu_{\,\!}{\rm Al}\to e_{\,\!}{\rm Al})=($0.1-9.0$)\times10^{-15}$,\,
which are within reach of Mu2E and COMET, expected to have sensitivity levels
under $10^{-16}$ or better after several years of running\,\,\cite{CeiA:2014wea}.

\end{document}